# Extending the Black-Scholes Option Pricing Theory to Account for an Option Market Maker's Funding Costs

Lou Wujiang[1]
HSBC,
March 8, 2014, updated Aug 22, 2014

**Abstract**

An option market maker incurs funding costs when carrying and hedging inventory. To hedge a net long delta inventory, for example, she pays a fee to borrow stock from the securities lending market. Because of haircuts, she posts additional cash margin to the lender which needs to be financed at her unsecured debt rate. This paper incorporates funding asymmetry (borrowed cash and invested cash earning different interest rates) and realistic stock financing cost into the classic option pricing theory. It is shown that an option position can be dynamically replicated and self financed in the presence of these funding costs. Noting that the funding amounts and costs are different for long and short positions, we extend Black-Scholes partial differential equations (PDE) per position side. The PDE's nonlinear funding cost terms create a free funding boundary and would result in the bid price for a long position on an option lower than the ask price for a short position. An iterative Crank-Nicholson finite difference method is developed to compute European and American vanilla option prices. Numerical results show that reasonable funding cost parameters can easily produce same magnitude of bid/ask spread of less liquid, longer term options as observed in the market place. Portfolio level pricing examples show the netting effect of hedges, which could moderate impact of funding costs.

## 1. Introduction

The Black-Scholes (BS) option pricing theory [1, 2] has become the foundation of modern finance, in both theory and practice. A number of extensions have been developed since, either to relax assumptions or to deal with practical issues, for example, commodity options [3], American options on dividend paying stocks [4], and implied volatility smile [5] via local or stochastic volatility [6, 7].

Concerning funding assumptions, although the text book presentation of the theory references to the risk free rate [8, 9], there has been some latitude in practice in terms of which rate(s) to use as a proxy of the risk free rate. Prior to the financial crisis, LIBOR has been generally accepted as the risk free rate in use of option pricing models. A trader buying and selling stock to delta hedge an option portfolio readily finds that her stock borrowing or repo rates are more relevant. Implied repo rate, for example, is popular in the equity forward market where the forward price is driven by the risk-free rate, the stock dividend yield and the stock repo rate (or spread). A seemingly intuitive workaround is to simply use the repo rate as the stock's rate of return in the Black-Scholes equation.

---

[1] The views and opinions expressed herein are the views and opinions of the author, and do not necessarily reflect those of his employer and any of its affiliates.

Following the financial crisis, overnight indexed swap (OIS) curve has taken the place of the LIBOR curve as the risk free discount curve in fully collateralized or exchanged traded derivative products pricing. The OIS curve, however, is a cash deposit curve in nature while the LIBOR curve is a benchmark borrowing curve. Cash deposit and borrowing of cash are asymmetric in that they carry a spread or interest margin. In developed economies, interest margin is small and at a minimum most of the times, although liquidity stress could produce large deviations from the norm, for instance, the 400+ basis point jump in the three month LIBOR and OIS spread during the summer of 2008. In emerging economies, interest margin can be regulated and excessive[2]. Lou [2010] introduced a debt account into the option economy that pays the bank's senior unsecured rate on its cash borrowing. The derived BSM equation has a term reflecting the asymmetric funding cost as the product of the asymmetric funding spread and the debt account balance, which can be defined to maintain a perfect dynamic replication of the option. Option price naturally deviates away from the classic BSM formulae and the difference is termed as a funding valuation adjustment (FVA).

Funding costs also come from haircut, a standard market mechanism in securities lending markets. Effectively an overcollateralization, haircut creates a funding deficiency that has to be filled with the borrower's unsecured cash borrowing. For investment grade corporate bonds, haircuts could range from 8% to 20%. For stocks, haircuts could range 15% to 50%. In Basel II and III the standard supervisory haircut for an equity appearing in a main index is 25%. As is with the funding rate, levels of haircuts depend on the borrower's credit and asset quality, and can be quite dynamic. Gorton and Metrick [14] show that haircuts for certain asset sectors rose up to 50% during the financial crisis. While modeling haircut could be tedious, it is unrealistic not to do.

An option market maker may have to carry net option positions which could be delta hedged until the inventory is cleared. In doing so, she incurs funding cost. Option market making literature has focused on liquidity, order flow, utility, discrete time hedge, unhedgeable risk such as stochastic volatility [15]. An interesting question is how funding cost contributes to option price bid and ask spread. In this paper we focus on options traded on well established exchanges where option trading accounts are guarded with proper margin requirements to an extent such that counterparty default risk can be ignored. By carefully examining the funding needs and solutions, we derive different PDEs for a long option position and a short option position. The respective solutions of these long and short PDEs then produce a bid (long) and ask (short) spread.

**2. Long and Short Position PDEs with Funding Cost**

We start with a brief recap of the funding cost modeling framework in Lou [10,11]. Consider a simple option economy where an option market maker undertakes a dynamic trading strategy to hedge a European vanilla option, expiring at time $T$ and fair valued at $V_t$. The market maker trades a hedging strategy $\Delta_t$ in the underlying stock with price $S_t$. Trading is assumed to be continuous and frictionless.

---

[2] China, for example, has had for a long time a floor on lending rates and cap on deposit rates, attributing to a 3% of asymmetric funding spread, in the same magnitude as the deposit rate.

To finance options trading and hedging, funds from different sources can be arranged. The initial purchase money, for example, can come from the market maker's own capital, borrowed money, or asset generated cashflow. It is not unreasonable to assume that an active market maker's borrowing cost is lower than its return on equity, so the preference is given to borrowed money, especially secured borrowing. We assume that the amount of borrowing does not correlate with the borrowing cost.

## 2.1 Stock Financing Setup

If the market maker buys stock, she could borrow at a lower rate than unsecured borrowing because the stock can be used to secure the borrowing, although such a rate is still higher than the risk free rate. If she shorts stock, the proceeds from sales do not earn the deposit rate as the stock lender deducts a borrowing fee. Long or short, both involve costs. This contrasts with the standard assumption in the classical BSM framework where money borrowed to buy stock carries the risk free rate and short sales proceeds also earn the risk free rate.

Stock financing, whether in repo (to long) or sec lending (to short) form, involves overcollateralized margin accounts. In a repo trade, the cash borrower can only borrow a cash amount at a discount to the market value of stock. Let $S_t$ be the stock price, $h$ haircut, the amount to borrow against one share of stock is $(1-h)S_t$. As the haircut amount $hS_t$ needs to be financed separately, often at the market maker's senior unsecured rate, additional funding cost occurs.

To short stock, shares have to be borrowed from the securities lending market and cash proceeds from the stock sales are posted to a cash margin account as a lien, which alleviates the stock lender's potential loss should the stock price rise and the borrower default. The margin account is in fact overcollateralized as well. Let $h^*$ be haircut for sec lending, the required total cash margin is $(1+h^*)S_t$, of which the additional cash of $h^*S_t$ needs to be financed.

The stock borrower typically receives interest on the total amount of cash posted at a rebate rate lower than the return the lender earns on investing the cash. For simplicity and modeling consistency, we consider cash deposit as the only investment vehicle. Let $r_p$ be the interest rebate rate, $r_p < r$ so that $r - r_p$ is the stock's borrowing cost. The borrower pays a manufactured dividend to the lender as it happens. The margin account is revaluated continuously and cash margin moves accordingly. Theoretically, one can also look at a reverse repo transaction to borrow and sell stocks, although shorting through the reverse repo market is mostly for the treasuries. The sec lending mechanism described above, however, adequately covers a reverse repo.

As is standard of a secured financing, the stock or cash as collateral are transferred to the other party so that the pledging party no longer has the ability to use the collateral in any other ways. This rules out the possibility of lending out stock purchased with a repo finance to earn sec lending fee which would then offset the fee paid on the repo.

To summarize, we add a repo account to the option economy, with balance of $R_t$, repo rate $r_p(t)$. If $R_t > 0$, the market maker finances its stock purchase with a repo trade, otherwise, she borrows shares from a sec lender. A repo account balance is tied up to $\Delta$ shares lent or borrowed, $R_t = (1-h)\Delta S_t$, where $h>0$ for a repo trade and $h<0$ for a sec lending trade.

## 2.2 Short Option PDE

Consider the market maker with a short option which is dynamically hedged in an economy that has access to stock financing markets. The market maker maintains a bank account with balance $M_t \geq 0$, which earns interest at the risk-free rate $r(t)$. Its unsecured cash borrowing is captured in the debt account balance $N_t \geq 0$, paying rate $r_b(t)$. The cash and debt accounts are short term or instantaneous in nature, each having a unit price of par, and rolled at the short rates $r(t)$, $r_b(t)$ respectively, $r_b(t) \geq r(t)$.

On the asset side of the balance sheet, the economy has the deposit account and stock; on the liability side, option valued at $V_t$, repo borrowing $R_t$, and unsecured borrowing $N_t$. The wealth of the economy $\pi_t$ is simply,

$$\pi_t = M_t + \Delta_t S_t - V_t - R_t - N_t,$$

Over a small time interval $dt$, the economy will rebalance the hedge and financing. Assuming that $d\Delta$ shares of stock are purchased at the price of $S_t + dS_t$ and subsequently added to the repo account, there will be a cash outflow of $d\Delta_t(S_t + dS_t)$ amount and inflow of $dR_t$ resulting from increased repo financing. The repo account arrives at a new balance, and collects an interest amount $r_p R_t dt$. The economy receives stock dividend of $q\Delta S_t dt$ where q is the stock dividend yield. The debt account pays interest amount $r_b N_t dt$ and rolls into new issuance amount of $N_t + dN_t$. The cash account accrues interest amount $rM_t dt$. Any net positive cash flow is deposited into the bank account, while net negative cash is borrowed in the unsecured account.

These cash flow activities are closed in that there is no capital inflow or outflow and that all investment activities are fully funded within the economy. In other word, the economy is self-financed. Summing up, we have the financing equation,

$$dM_t = rM_t dt - d\Delta_t(S_t + dS_t) + \Delta_t S q dt + dN_t - r_b N_t dt + dR_t - r_p R_t dt$$

Differentiate the wealth equation and plug in the financing equation,

$$d\pi_t - r\pi_t dt = \Delta_t(dS_t - (r-q)S_t dt) - (dV_t - rV_t dt) - (r_b - r)N_t dt - (r_p - r)R_t dt$$

Incrementally, the excess return of the net worth of the economy (on the left hand side) equals the net excess return on the stock and the option, minus the financing cost of the repo and the unsecured debt. Under a diffusion dynamics, $dS = S(\mu dt + \sigma dW)$, choose a delta hedge strategy $\Delta_t = \frac{\partial V_t}{\partial S}$, note $R_t = (1-h)\Delta_t S_t$, and apply Ito's Lemma to obtain the following,

$$d\pi - r\pi dt = -\left(\frac{\partial V}{\partial t} + (r_s - q)S\frac{\partial V}{\partial S} + \tfrac{1}{2}\sigma^2 S^2 \frac{\partial^2 V}{\partial S^2} - rV + (r_b - r)N\right)dt$$

where $r_s = r + (1-h)(r_p - r)$ can be seen as a nominal stock rate of return. In order to fully replicate the option, we set $\pi_t = 0$ for $t \geq 0$, i.e., the portfolio starts out at zero and maintains zero at all times. From the wealth equation, $M_t - N_t = V_t - h\Delta_t S_t$. By choosing the unidirectional funding strategy ($M_t \cdot N_t = 0$ for $0 \leq t \leq T$), the funding strategy pair ($M_t$, $N_t$) becomes defined,

$$M_t = (V_t - h\Delta_t S_t)^+,$$
$$N_t = (h\Delta_t S_t - V_t)^+$$

The deposit account balance is the value of the derivative subtracted by the amount used to fund the repo margin. Alternatively, the unsecured borrowing amount is the repo margin amount subtracted by the derivative value. Finally from $d\pi_t=0$, we arrive at the extended Black-Scholes PDE for a short option position,

$$\frac{\partial V}{\partial t} + (r_s - q)S\frac{\partial V}{\partial S} + \tfrac{1}{2}\sigma^2 S^2 \frac{\partial^2 V}{\partial S^2} - rV + (r_b - r)(hS\frac{\partial V}{\partial S} - V)^+ = 0$$

Here the option is fully hedged and the funding strategy is well defined. The PDE is however no longer linear and we can not treat $r_s$ as the stock return under the risk neutral measure. In general, when stock is financed through the repo market at a non-zero haircut, the notion of a risk neutral stock return needs to be carefully examined.

The classic BSM equation can be derived in two equivalent approaches, the dynamic replication approach originally by Black and Scholes [1], and Merton [2], and risk neutral pricing by Harrison and Kreps [15], and Harrison and Pliska [16]. By following the same dynamic replication setup, we can take out the assumption of a risk-free borrowing and yet arrive at a no-arbitrage solution. As is in the classic BSM theory, a perfect delta hedging strategy and accompanying self funding strategy exist, no hedging error is resulted and thus no risk capital is required. Risk neutral pricing however is based on the assumption of deposit and borrowing at the same risk free rate which every investment in the risk neutral world would earn in order to guarantee no arbitrage. It will be interesting to see how a multi-curve setup, e.g., different deposit and borrow curves, can be built into the risk neutral pricing framework.

**2.3 FVA and Option Market Making**

First of all, the PDE for a short option is different from the PDE for a long position ( Lou[11] ),

$$\frac{\partial V}{\partial t} + (r_s - q)S\frac{\partial V}{\partial S} + \tfrac{1}{2}\sigma^2 S^2 \frac{\partial^2 V}{\partial S^2} - rV - (r_b - r)(V - hS\frac{\partial V}{\partial S})^+ = 0$$

Specifically, the last term is a cost (negative sign in front), while for a short option, it has a positive sign. If we replace V with −V in the (long) PDE above, then the short PDE is immediately recovered. Noticing that the position value of a long option priced at V is V and a short position −V, the long PDE in fact applies to both long and short positions if V is treated as an option position value rather than price of the option.

The PDE's last term indicates that we need to pay for its unsecured cost, whether long or short an option, a direct consequence of the funding asymmetry. Long and short positions however entail different unsecured funding amounts. To long a call option, for example, requires to pay for the option purchase and to post the additional cash margin. To short the call needs to post additional cash repo margin as the option is paid for. Subject to funding cost adjustment, a long position becomes less valuable than the risk free rate funded option, and a short position more expensive. Consequently a price difference between a long and a short position is generated and will contribute to option market makers' bid and ask spread. Let $V^*$ be the usual B-S fair price without funding cost adjustment, $V_b$ and $V_a$ be the bid (long) and ask (short) prices of the option market maker. If we denote $f_b$ as FVA for a long position, and $f_a$ short position, then it follows by definition, $f_b = V^* - V_b, f_a = V_a - V^*$.

For vanilla options, it is possible to simplify the extended BSM equation, as shown below.

### 2.3.1 Long vanilla call and put options

For a long call option, price $V$ and delta $\Delta$ are both nonnegative. Hedging would require shorting stocks borrowed from the stock lending market with a haircut $h<0$, so that we have $(V - hS\frac{\partial V}{\partial S})^+ = (V - hS\frac{\partial V}{\partial S})$, and the long PDE becomes,

$$\frac{\partial V}{\partial t} + (hr_b + (1-h)r_p - q)S\frac{\partial V}{\partial S} + \tfrac{1}{2}\sigma^2 S^2 \frac{\partial^2 V}{\partial S^2} - r_b V = 0$$

Intuitively one needs to finance the sum of the option and the residual margin. This applies to a long put option as well, where $\frac{\partial V}{\partial S} \leq 0$ and $h \geq 0$ as to hedge the long put we need to long stocks by utilizing repo financing. The specifics of repo or sec lending can be seen if we rewrite the PDE as follows,

$$\frac{\partial V}{\partial t} + \tfrac{1}{2}\sigma^2 S^2 \frac{\partial^2 V}{\partial S^2} - qS\frac{\partial V}{\partial S} + (1-h)r_p S\frac{\partial V}{\partial S} - r_b(V + S\left|h\frac{\partial V}{\partial S}\right|) = 0$$

The fourth term reflects the repo financing cost while the last term shows the unsecured borrowing cost. For short options, one needs to finance the difference of the option and additional margin, which is not guaranteed to be positive or negative.

### 2.3.2 Zero haircut call and put spread

Let $h=0$, $r_1$ be the repo rate, $r_1>r$, $r_1-r$ the repo spread, and $r_2$ the sec lending rebate rate, $r_2<r$, $r-r_2$ the stock borrowing cost. Short call and long call equations degenerate respectively to,

$$\frac{\partial V_a}{\partial t} + (r_1 - q)S\frac{\partial V_a}{\partial S} + \tfrac{1}{2}\sigma^2 S^2 \frac{\partial^2 V_a}{\partial S^2} - rV_a = 0,$$

$$\frac{\partial V_b}{\partial t} + (r_2 - q)S\frac{\partial V_b}{\partial S} + \tfrac{1}{2}\sigma^2 S^2 \frac{\partial^2 V_b}{\partial S^2} - r_b V_b = 0$$

The long option PDE shows that the market maker finances the purchase with money raised from the unsecured market, and the short position PDE shows that she receives cash from the buyer and cash earns the risk free rate. Obviously this is in a context of no counterparty credit risk. For European options, analytical formula exists, see for example Shreve [8]. Assuming all rates are flat, for example, we have a bid-ask spread of a call option with strike K in analytical form,

$$V_a - V_b = e^{-r(T-t)}[F_a \phi(d_1(F_a)) - K\phi(d_2(F_a))] - e^{-r_b(T-t)}[F_b \phi(d_1(F_b)) - K\phi(d_2(F_b))],$$

$$F_a = Se^{(r_1-q)(T-t)}, F_b = Se^{(r_2-q)(T-t)},$$

$$d_1(F) = \frac{1}{\sigma\sqrt{T-t}}[\ln(\frac{F}{K}) + \tfrac{1}{2}\sigma^2(T-t)], d_2(F) = d_1(F) - \sigma\sqrt{T-t}.$$

where ϕ() is the standard normal cumulative distribution function.

For a put, delta is negative. A long put position requires long stock to hedge and repo financing applies. A short put position needs to hedge with short stock. Short put and long put equations are the same as the call option PDEs except the rates $r_1$ and $r_2$ are interchanged to reflect the opposite use of the stock financing.

## 3. Finite Difference Solution and Numerical Examples

### 3.1 Finite Difference Solver

With the one-sided unsecured funding charge term in the extended PDE, a numerical solution has to be sought. Finite difference (FD) methods for European and American vanilla options are well studied [17]. The nonlinear term in our PDE remains continuous and is practically differentiable everywhere except at some point where the funding account kicks in or extinguishes. In fact, *(t,S)* domain of the PDE can be subdivided into two regions where one has $N_t>0$ while the other has $N_t=0$, where each region then has a B-S type PDE with differing stock and derivative financing rates. The free boundary separating these two regions can be sought with a simple iterative procedure. Across the free boundary, we expect continuity of option price and delta.

For a single option position, whether European or American, delta is one-sided, i.e., either positive or negative. The nominal stock financing rate (the coefficient of the $\frac{\partial V}{\partial S}$ term) is a constant, given fixed haircut and repo rate. For more complex option trading strategies, being net long delta or net short delta depends on local stock price. The nominal stock financing rate is localized as the net delta sign decides whether a repo financing or a sec lending charge applies.

We developed a Crank-Nicholson scheme to solve the PDE for European style options, with an iterative procedure to solve for the funding boundary. For American options, the standard PSOR (projected successive over relaxation) method is modified to iterate on the funding term as well as early exercise boundary. To allow for pricing of an option trading strategy or combination (such as straddle, strips), we adopt zero gamma boundary conditions on both upper and lower boundaries. The PDE is applied at the half node nearest to the boundary so that the resulting set of linear difference equations remains tri-diagonal. The FD solver is calibrated to the B-S formula, see Table 1 for comparisons with analytical results.

Table 1. Comparison of FD solver results with B-S analytic results, European put&call, S=K=100, T=2, r=0.1, vol=0.5. FD: dt=0.02, S grid of 2000 nodes.

| Call | Analytic | FD | Put | Analytic | FD |
|---|---|---|---|---|---|
| Price | 35.1452 | 35.1445 | Price | 17.0183 | 17.017279 |
| Delta | 0.737741 | 0.737782 | Delta | -0.262259 | -0.262254 |
| Gamma | 0.0046077 | 0.0046522 | Gamma | 0.0046077 | 0.0046523 |

**3.2 Single Option Pricing with Funding Costs**

As shown earlier, the fair price of a long position in vanilla European options has analytic formula. Figure 1 plots the FVA percentage of an at-the-money (ATM) put as a function of the market marker's unsecured funding spread $r_b$-$r$, when repo financing is not used (curve legend "no-repo", equivalent to $h=1$ or $h=-1$), and when repo is utilized under three combinations of repo haircuts and repo rates, namely zero haircut with repo rate of 50 basis points, 35% haircut with 50 bps, and 35% haircut with 150 bps. The risk-free B-S price of the put is 17.0183, with parameters S=100, K=100, vol = 50%, T = 2 years, r =10%, q =0. Under each of the four cases, FVA increases as the firm's spread increases from a shown range of zero to 4%. With haircut fixed at 35%, higher repo rate (150 bps compared versus 50 bps) results in higher FVA. With repo rates fixed at 50 bps, the FVA curves under zero and 35% haircuts show a crossover at 0.5% spread, which can be easily explained from the long put PDE that when the repo rate and unsecured spread are the same, the PDE does not depend on haircuts.

To get some sense of real world option bid/ask spread levels, Table 2 lists Jan 2015 expiry puts and calls on Goldman Sachs' stock as of July 31, 2013. Mid market prices and spreads are shown for ATM strike of 165 and two nearest in and out strikes. The call and put spreads are sizeable (up to 3.65 points) and range from 10% to 20% of the mid option prices. The last two columns show preliminary results of call and put bid/ask spreads of similar magnitudes, calculated assuming the same implied volatility and dividend yield, and the market maker's spread of 300 bps, repo spread 70 bps, and haircut 25%.

Obviously funding cost is not the only factor affecting option bid/ask spread, but its contribution to long dated option can not be ignored. Stoikov and Saglam [14] studied market making under option inventory risk. If the market maker carries a net inventory, end-of-day option hedging positions are needed, and the funding cost will incur. Cetin et al [18] showed liquidity's impact on market making.

Table 2. Sample long dated option bid/ask spreads

| GS Option | Market - July 2013 | | | | Calculated | |
|---|---|---|---|---|---|---|
| Strike | Mid Call | Mid Put | Call Sprd | Put Sprd | Call Sprd | Put Sprd |
| 155 | 23.6 | 15.95 | 3.5 | 3.2 | 2.98 | 2.48 |
| 160 | 20 | 18.125 | 3 | 2.85 | 2.88 | 2.74 |
| 165 | 17.625 | 20.475 | 1.75 | 3.65 | 2.74 | 3.01 |
| 170 | 15.375 | 23.075 | 2.85 | 3.65 | 2.59 | 3.28 |
| 175 | 13.2 | 26.05 | 2.8 | 3.7 | 2.42 | 3.54 |

### 3.3 Option Portfolio Results

A typical option market maker would make market in an option chain, i.e., a series of options of various strikes and different expiries on the same underlying stock. As a result, the market maker only needs to finance the net stock position, whether long or short. Consequently funding cost's impact is expected to moderate. To illustrate, we consider some basic option trading strategies and compute FVA with netting effect as compared to a single option where no netting effect exists.

Figure 2 shows the bid/ask spread of a bull spread where the market maker is long a call at 95 strike and simultaneously short a call at 105 strike. The backward finite difference solver for the extended B-S PDE can easily accommodate the strategy's terminal payoff and the zero gamma boundary condition is not concerned with the details of the payoff. The solid line plots the bull bid/ask spread as a function of expiry, showing mostly small values with an increasing trend due to funding cost accumulation. The dashed "Synthetic" curve shows the synthetic portfolio of C95 – C105 where C95 and -C105 are separately priced long 95 call and short 105 call positions. The netting effect is defined as the difference between the bull spread and the "Synthetic" bull spread. As one position is long delta while the other position is short delta in the same magnitude, the netting effect is very large.

A straddle is a volatility trading strategy where a party buys a call and a put of the same strike and expiry [9]. Since long a call has positive delta and long a put has negative delta, netting of delta takes effect. Figure 3 shows the straddle's bid/ask spread computed with and without (labeled as Synthetic) netting effect. The funding cost induced bid/ask spread is much more pronounced than a bull spread as seen in Figure 2.

Similar to a straddle, a strangle buys a call and a put with the same expiry date, but with the call having higher strike than the put. Figure 4 shows the bid/ask spread of such a strangle where the call strike is 105 while the put is 95.

A strip is a tilted combination where the party buys one call and two same strike and expiry puts. Netting of positive and negative deltas are less balanced than that of a straddle and as a result the bid/ask spread due to funding cost accumulation is higher, as seen in Figure 5.

### 4. Concluding Remarks

This paper considers funding costs faced by option market makers and impact on option valuation. The Black-Scholes option pricing theory is extended to admit separate rates for cash deposit and borrowing, in a no-arbitrage, fully replicated manner. Deposit or borrowing amounts capture realistic funding of the option and the underlying stock hedge, including overcollateralization (haircut) requirement prevalent in the stock financing markets. The gap or asymmetry between the deposit and borrowing rates alone is significant for emerging markets where a structural deposit and borrowing rate disparity exists. We find that funding of long position and short position is inherently asymmetric, i.e., funding cost adjusted option bid price is lower than ask price, naturally creating a bid/ask spread.

The extended Black-Scholes PDE has a non-linear funding cost adjustment term attributed to a free funding boundary which makes it necessary to seek numerical solutions, except for long options. For a short position, traditional finite difference scheme needs to be modified to define the funding boundary. Numerical results have shown to produce a bid-ask spread in the same magnitude as observed in the options market. Options book level netting effect is shown to be significant, mostly due to the netting of deltas.

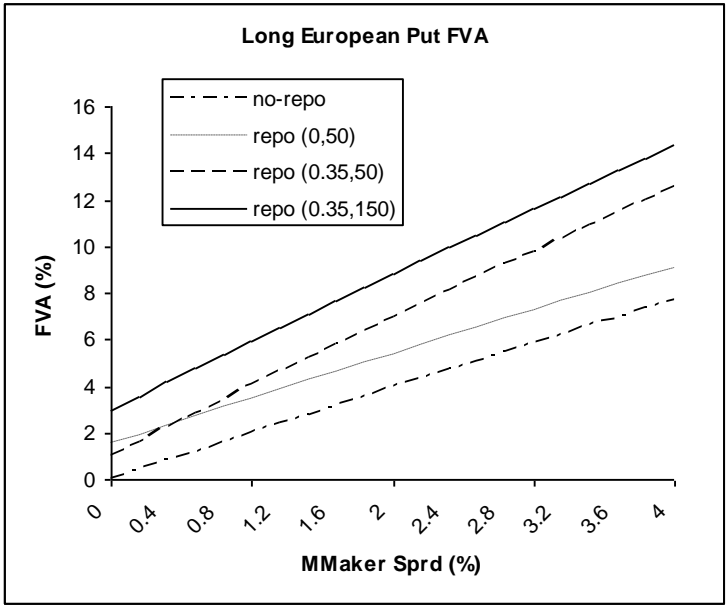

Figure 1. FVA percentage of B-S fair price of an ATM long European put option, S=K=100, vol=50%, T=2 years, r =10%.

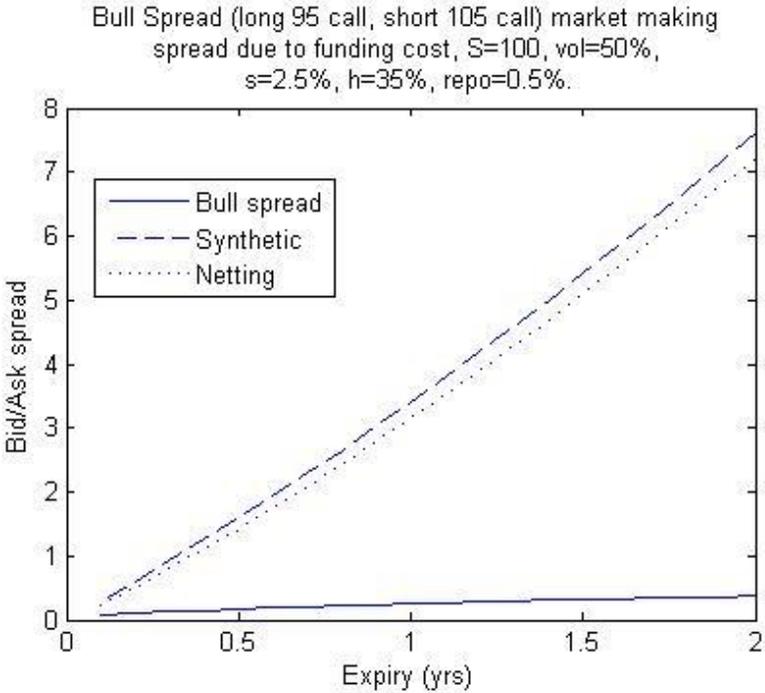

Figure 2. A bull spread trading strategy's funding induced bid/ask spread and netting effect.

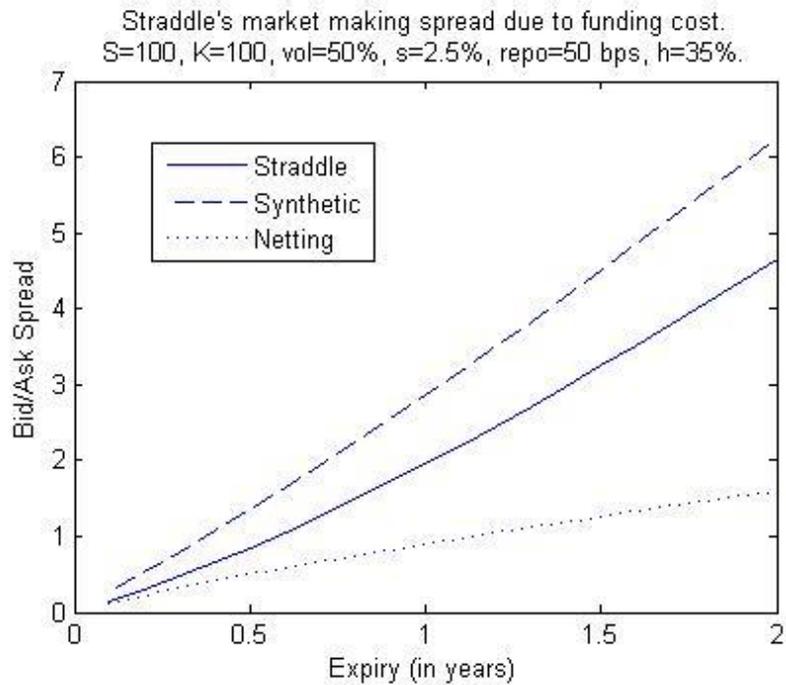

Figure 3. A straddle trading strategy's funding induced bid/ask spread and netting effect.

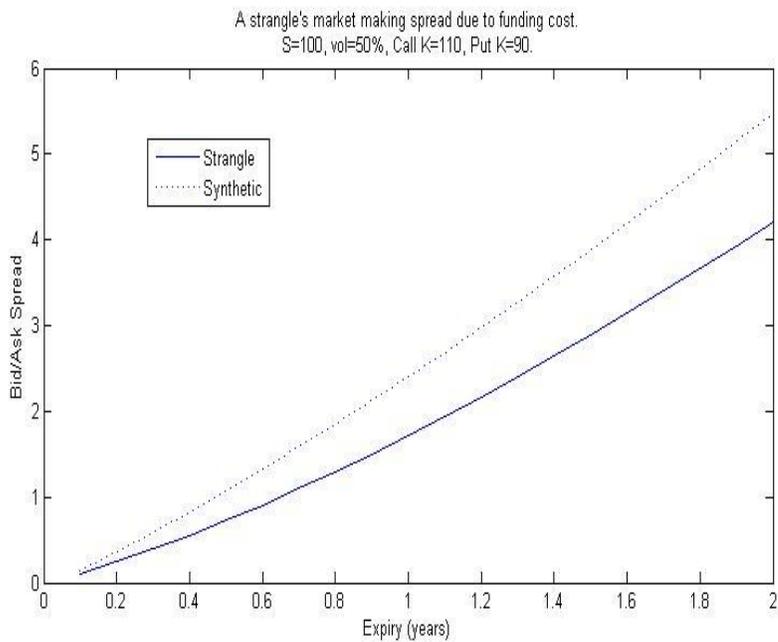

Figure 4. A strangle trading strategy's funding induced bid/ask spread and netting effect.

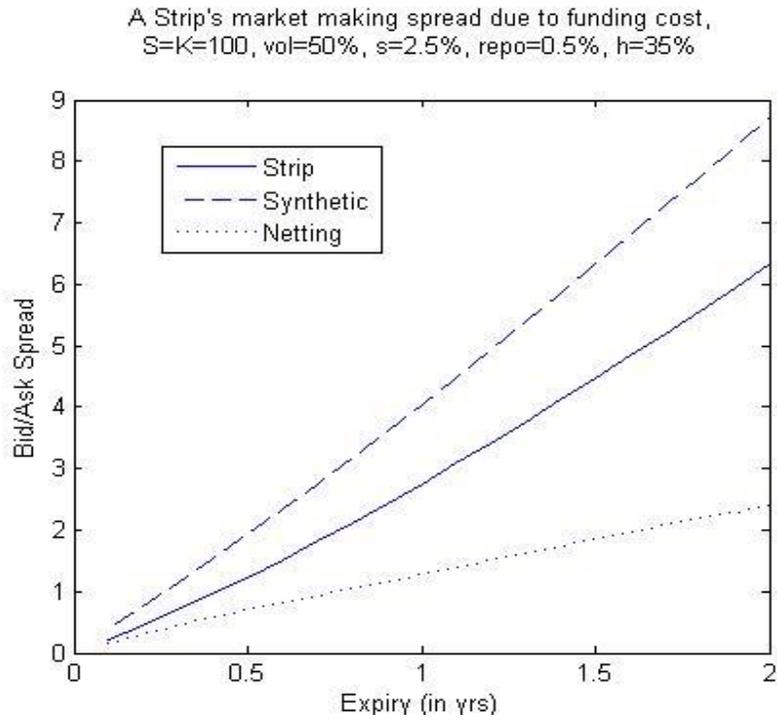

Figure 5. An option strip trading strategy's funding induced bid/ask spread and netting effect.